\documentclass[final,authoryear,5p]{elsarticle}

\usepackage{graphicx}
\usepackage{multirow}

\usepackage{amssymb}
\usepackage{amsmath}

\usepackage[
a4paper=true,%
breaklinks=true,%
colorlinks=true,%
pdfauthor={Boschini et al.},%
pdftitle={Propagation of Cosmic Rays in Heliosphere: the HelMod Model}%
]{hyperref}

\journal{Advances in Space Research}

\biboptions{authoryear}

\newcommand{\degree}{\ensuremath{^\circ }}
\newcommand{\galprop}{GALPROP}
\newcommand{\helmod}{\textsc{HelMod}}

\begin{document}

\begin{frontmatter}


\title{The \helmod{} Model in the Works for Inner and Outer Heliosphere: from AMS to Voyager Probes Observations.}
\author[label1,label2]{M.~J.~Boschini}
\author[label1]{S.~{Della~Torre}}
\author[label1,label3]{M.~Gervasi}
\author[label1,label3]{G.~{La~Vacca}}
\author[label1]{P.~G.~Rancoita\corref{cor}}
\cortext[cor]{Corresponding author}
\ead{piergiorgio.rancoita@mib.infn.it}
\address[label1]{INFN sez. Milano-Bicocca, Piazza della Scienza, 3 - 20126 Milano (Italy)}
\address[label2]{also CINECA, Segrate, Milano, Italy}
\address[label3]{also Physics Department, University of Milano-Bicocca, Piazza della Scienza, 3 - 20126 Milano (Italy)}

\begin{abstract}
\helmod{} is a Monte Carlo code developed to describe the transport of Galactic Cosmic Rays (GCRs) through the heliosphere from the interstellar space to the Earth.
In the current \helmod{} version 4 the modulation process, based on Parker's equation, is applied to the propagation of GCRs in the inner and outer heliosphere, i.e., including the heliosheath.
\helmod{} was proved to reproduce protons, nuclei and electrons cosmic rays spectra observed during solar cycles 23-24 by several detectors, for instance, PAMELA, BESS and AMS-02. In particular, the unprecedented accuracy of AMS-02 observations allowed one a better tuning of the description regarding the solar modulation mechanisms implemented in \helmod{}. In addition, \helmod{} demonstrated to be capable of reproducing the fluxes observed by the Voyager probes in the inner and outer regions of heliosphere up to its border.
\end{abstract}


\end{frontmatter}

\parindent=0.5 cm

\section{Introduction}\label{Introduction}
The increased performances of space-borne spectrometers enhanced
the accuracy of observed modulated omni-directional Galactic Cosmic Rays (GCRs) spectral intensity. Furthermore, Voyager probes provided the first and direct observations of Termination Shock and Heliopause. This effort leads to a better capability a) to unveil local interstellar spectra (LIS) of GCR species~\citep[e.g., see][and references therein]{2017ApJ...840..115B,2018ApJCO,2018ApJElectron}, b) to investigate their generation, acceleration and diffusion process within the Milky Way \citep[e.g., see][]{Boella1998,StrongMoskalenko2007,EvoliGAggero2008,PutzeMaurin2009}, and, in turn, c) to possibly untangle features related to new physics -- i.e.,~due to dark matter \citep[e.g., see][and references therein]{BottinoDonato1998,CirelliCline2010,IbarraTran2010,Salati2011,Weniger2011} --  or to additional astrophysical sources so far not taken into account \citep[e.g., see][and references therein]{Chang2008, AbdoetAl2009,AdrianietAl2009a,Cernuda2011,MertschSarkar2011,RozzaJHEA_2015,ICRC2015_Rozza}. 

A deep understanding of the solar modulation effect -- i.e. the physical process affecting GCRs transport in the heliosphere -- 
is needed for the investigations described above. It can be obtained by systematic studies of GCRs spectra observed during different phases of solar activity by experiments operated on stratospheric balloons~\citep[for instance, see][]{1983ApJ...275..391W,BoezioetAl1999,Mennetal2000,HainoSanuki2004,bess_prot,AbeetAl2008,BESS2007_Abe_2016} or in space-borne missions~\citep[e.g., see][and reference therein]{1983ApJ...275..391W,1995SoPh..162..483M,AMS_protons,AMS_leptons,AMS_cosmic,AMS_helium,AMS01_prot,AMS_positron,AdrianietAl2009a,AdrianietAl2009b,AdrianietAl2010,PAMELA_Prot_He_2011,PamelaProt2013,AMS02_2014_PhysRevLett2,AMS02_2015_PhysRevLett2,AMS02_2015_PhysRevLett1,AMS02_2016_PhysRevLett1,AMS02_2016_PhysRevLett2,2017PhRvL.119y1101A,2018PhRvL.pHe,2018PhRvL.121e1103A,2018PhRvL.120b1101A,2018PhRvL.epos,2017JGRA..122.1463K}.
\par
Among space missions currently detecting GCRs,
AMS-02 -- on-board of the International Space Station since May 2011 -- observed GCRs flux during the maximum of solar cycle 24, providing data with unprecedented measurement accuracy~\citep{2018PhRvL.pHe,2018PhRvL.epos}.
On the other hand, Ulysses made unique measurements in the inner part of heliosphere observing particle radiation outside the ecliptic plane~\citep[e.g., see][]{Simpson1992,Simpson1996b,Heber1996,FerrandoEtAl1996}. 
Among its achievements, it reveals the presence of latitudinal particle intensity gradients whose magnitude depends on solar magnetic field polarity~\citep{deSimone2011,GieselerHeber2016}.
Besides, Voyager probes explored heliosphere up to its boundary (and beyond) and showed the truly un-modulated Local Interstellar Spectrum (LIS) of GCRs~\citep{1986GeoRL..13..781M,CummingsetAl1987,1990SSRv...52..121V,2005IJMPA..20.6727Z,Krimigis144,Decker2020,Decker2008,Stone2017,Stone150,Webber2013}.
High precision AMS-02 experimental spectra, together with observations from Ulysses spacecraft outside the ecliptic plane and from Voyager probes up to the heliosphere boundary, represent a challenge for any modulation model. 
Finally, it is worth to remark that these data may, in addition, allow a better understanding of space radiation environment close to Earth, thus extending our capability to predict radiation hazards for astronauts and device damages in space missions~\citep[e.g., see][and Chapters 7 and 8 of~\citealt{rancoita2015}]
{leroyRancoita2007,ICRCDamage2015}.

\par
In this work we present the version 4.0 of HELiospheric MODulation (\helmod{}) model currently employed to solve the transport-equation for  GCR propagation through the heliosphere down to Earth~\citep{GervasiEtAl1999,Bobik2011ApJ,DellaTorre2013AdvAstro,BoschiniEtAl2017AdvSR}.
With respect to the previous code version~\citep{BoschiniEtAl2017AdvSR}, the present model improved the accuracy of particle transport mechanisms during solar maxima as described in Sect.~\ref{Sect:Model} and \ref{Sect::Results}.
The inclusion of a time dependent heliosphere boundary and of the heliosheath region (Sect.~\ref{Sect::OuterHeliosphere}) results in an enhanced modulation occurring at low energies and, thus, it required a re-tuning of modulation parameters as described in Sect.~\ref{Sect::Results}.

\begin{figure}
	\centering
	\includegraphics[width=1.\linewidth]{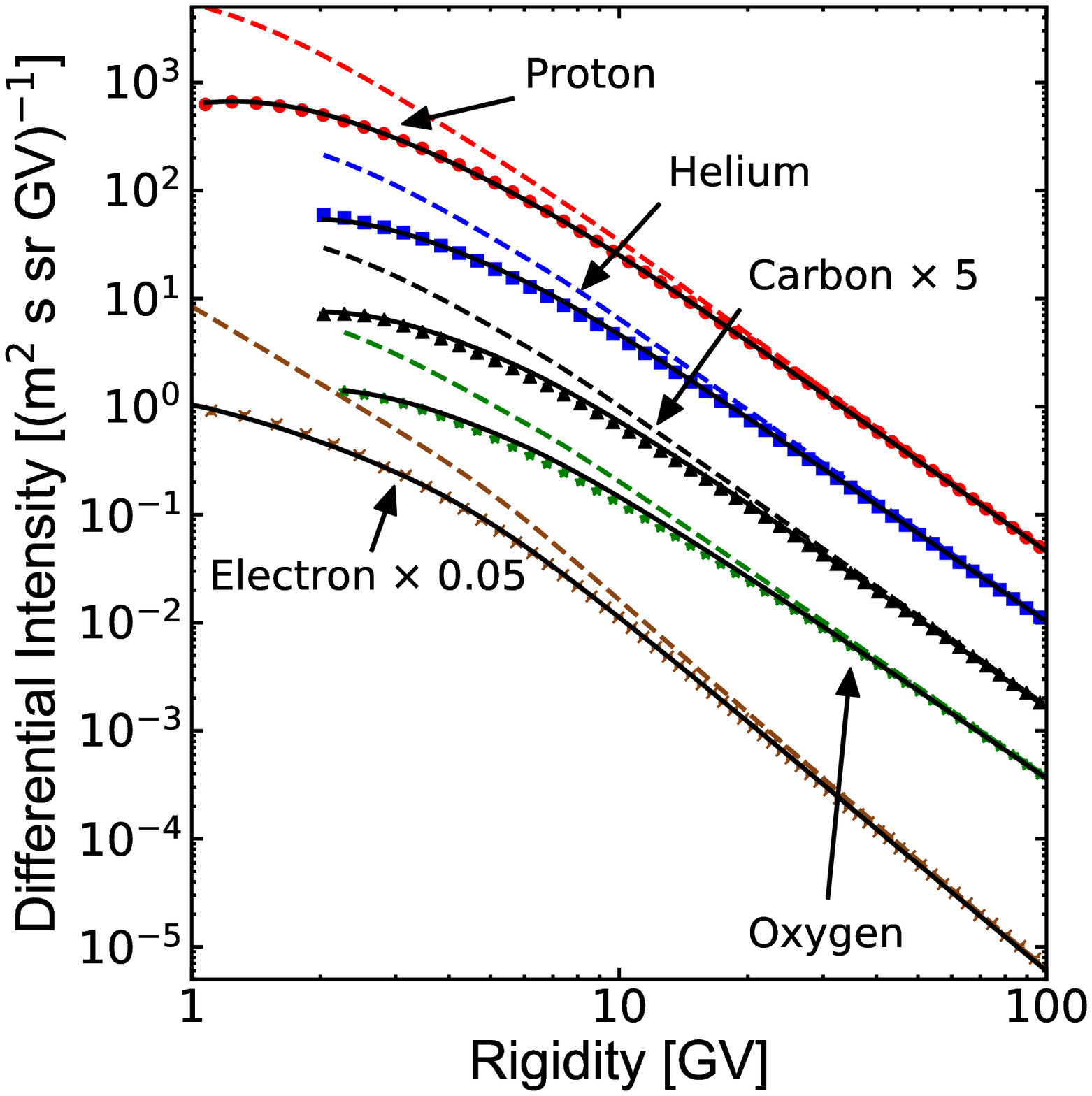}
	\caption{Summary of LISs (dashed line) obtained by means of
          \helmod{}-\galprop{} iterative procedure as reported in
          \citet[][for protons]{2017ApJ...840..115B},~\citet[][for
            helium, carbon and oxygen nuclei]{2018ApJCO},
          and~\citet[][for electrons]{2018ApJElectron}.
          The corresponding modulated spectra (solid lines)
          obtained using previous \helmod{} version 3 are shown with
          AMS-02 measurements (points) reported in \citet[for
            electron]{AMS02_2014_PhysRevLett2}, \citet[for
            Proton]{AMS02_2015_PhysRevLett1}, and \citet[Helium,
            Carbon and Oxygen]{2017PhRvL.119y1101A}.
	}
	\label{fig:helmodams02}
\end{figure}

\section{Heliospheric Propagation of Cosmic Rays}
\label{Sect:Model}
The particle transport through the heliosphere is a combination 
of several processes globally described  by the Parker Equation (\citealp{parker1965}, see, e.g., the discussion in 
\citealt{Bobik2011ApJ,BoschiniEtAl2017AdvSR} and reference therein):
\begin{align}
\label{EQ::FPE}
\frac{\partial U}{\partial t}= &\frac{\partial}{\partial x_i} \left( K^S_{ij}\frac{\partial \mathrm{U} }{\partial x_j}\right)\\
&+\frac{1}{3}\frac{\partial V_{ \mathrm{sw},i} }{\partial x_i} \frac{\partial }{\partial T}\left(\alpha_{\mathrm{rel} }T\mathrm{U} \right)
- \frac{\partial}{\partial x_i} [ (V_{ \mathrm{sw},i}+v_{d,i})\mathrm{U}],\nonumber
\end{align}
where $U$ is the number density of GCR particles per unit of kinetic energy $T$ (GeV/nucleon), $t$ is time, $V_{ \mathrm{sw},i}$ is the solar wind (SW) velocity along the axis $x_i$, $K^S_{ij}$ is the symmetric part of the diffusion tensor, $v_{d,i}$ is the particle magnetic drift velocity (related to the anti-symmetric part of the diffusion tensor), and $\alpha_{\mathrm{rel} }=\frac{T+2m_r c^2}{T+m_r c^2} $, with $m_r$ the particle rest mass per nucleon in units of GeV/nucleon.
The terms in the Parker Equation describe:
(i) the {diffusion} of GCRs scattered by magnetic irregularities,
(ii) the {adiabatic} energy losses/gains due to the propagation in the expanding magnetic fields carried in the SW,
(iii) an {effective convection} resulting from the SW convection with velocity $\vec{V}_{\rm sw}$,
and (iv) the magnetic {drift effects} related to the drift velocity ($\vec{v}_{\rm d}$).
Overall, the heliospheric modulation results in energy losses and suppression of the fluxes of CR species compared to the LIS. These effects are controlled by the level of solar activity, by the solar magnetic field polarity, and are energy- and charge-sign-dependent.

\par
It is widely accepted that $K^S$ components parallel to the magnetic field ($K_{||}$) are larger than perpendicular components ($K_{\perp,i}$), and should be described using non-linear theories \citep[for a review see, e.g.,][]{shalchi2009}.
At high rigidities (i.e., $\gg$1 GV) the diffusion tensor should have a linear (or quasi-linear)
rigidity dependence~\citep[e.g., see ][]{Gloeckler1966,1968ApJ...154.1011G,jokipii1966,jokipii1971,perko1987,potgieter1994,Strauss2011}.
The transition from the quasi-linear to the non-linear regimes results in a ``flattening'' of rigidity dependence at low values as observed, for instance, by~\citet{Palmer1982} and \citet{Bieberetal1994}.
In the present work 
we use a functional form with a rigidity dependence following the one presented in~\citet[and reference therein]{BoschiniEtAl2017AdvSR}:
\begin{equation}\label{EQ::KparActual}
K_{||}=\frac{\beta}{3} K_0\left( \frac{P}{1\text{GV}}+g_{\rm low}\right) \left(R_{c}+\frac{R}{\text{1 AU}}\right),
\end{equation}
where  $K_0$ is the diffusion parameter, which depends on the solar activity and magnetic polarity, $\beta$ is the particle speed in units of the speed of light, $P=qc/|Z|e$ is the particle rigidity in GV, $R$ is the heliocentric distance from the Sun in AU, and, finally, $g_{\rm low}$~\citep[discussed in ][]{BoschiniEtAl2017AdvSR} and $R_c$ are parameters  tuned to describe radial GCR intensity gradients on the inner heliosphere (see Sect.~\ref{Sect::Results}).

\par
The diffusion parameter $K_0$ fixes the normalization of $K_{||}$. It changes with time as defined by equation~(6--7) of \citet[][and references therein]{BoschiniEtAl2017AdvSR}. In turn, as introduced in~\citet{2018ApJElectron}, the diffusion parameter $K_0$ includes a correction factor that re-scales the absolute value of $K_{||}$ to account for the drift contribution.
This correction factor (discussed in Sect.~\ref{Sect::Results}) is evaluated using the proton flux during the period of positive polarity of the Heliospheric Magnetic Field (HMF) and applied to both electrons and ions when the condition $qA>0$ occurs; 
an additional correction has to be evaluated using electron flux and applied to negative-charged particle diffusion during positive HMF polarity period ($A>0$).

\par
The perpendicular diffusion coefficient is taken to be proportional to $K_{||}$ with a ratio
$K_{\perp,i}/K_{||}=\rho_i$ for both $R$ and $\theta$ $i$-coordinates~\citep[e.g., see ][and references therein]{potgieter2000, BurgerHattingh1998}.
At high rigidities, this description is consistent with quasi-linear theories (QLTs).
\citet{Palmer1982} constrains the value of $\rho_i$ between 0.02 and 0.08 at Earth.
In the current version of the model, we found $\rho_i\approx 0.065$ for protons and ions and $\rho_i\approx 0.05$ for electrons and positrons.
 The slight difference between $\rho_i$ values might be related to the mass differences between  massive particles (i.e, protons and nuclei) and leptons (i.e electron and positrons).
 As discussed in~\citet{Bobik2011ApJ}, we used an enhanced $K_{\perp,\theta}$ by a factor 2 in the polar regions in order to reproduce the amplitude and rigidity dependence of the latitudinal gradients of GCR differential intensities for protons~\citep[see Section 5.4 of][and reference therein]{BoschiniEtAl2017AdvSR}.


\par
As also remarked in~\citet{DellaTorre2013AdvAstro}, in this description $K_{||}$
has  a radial dependence $\propto R$ but no latitudinal dependence;
nevertheless, the reference frame transformations from the field-aligned frame to the spherical
heliocentric one~\citep[see, e.g.,][]{burg2008} introduce a dependence on the polar angle.
As was shown in \citet{DellaTorre2013AdvAstro}, this is enough to explain the
latitudinal gradient observed by Ulysses during the latitudinal \textit{fast scan}
in 1995~\citep[see e.g.][]{Heber1996,Simpson1996b}.

\par
We use the drift model originally developed by~\citet{Potgieter85} -- that includes description of \textit{regular drift} due to large scale structure of heliospheric magnetic field, and the \textit{neutral sheet drift} described, e.g, in \citet{1981ApJ...243.1115J,Hattingh1995} --,
and refined using Parker's magnetic field with polar corrections
as reported in~\citet{DellaTorre2013AdvAstro}~\citep[see also][for a discussion about modified Parker's magnetic field]{Raath2016}. 
Previous works underlined the importance of additional drift suppression during high activity periods~\citep[see, e.g., discussion in ][]{2004ApJ...603..744F,DellaTorre2013AdvAstro}.
This is due to the presence of turbulences in the interplanetary medium reducing the global effect of CR drift in the heliosphere~\citep{2017ApJ...841..107E}. 
In the present work we extend this description including a time dependent \textit{drift suppression factor} related to solar activity as discussed in Sect.~\ref{Sect::Results}.

LIS spectra are assumed to be nearly isotropic at the heliosphere boundary due to the relative small dimension of the heliosphere ($\sim$ 100 astronomical units) compared with the expected cosmic-ray density gradient scale in the galaxy ($\sim$ 18 pc) \citep[see, e.g., discussion in][]{2015PhPl...22i1501Z}. 
As described by~\cite{2017ApJ...840..115B}, in order to derive the physically motivated LIS of GCR species, an iterative procedure was developed to feed the \galprop{} output into \helmod{} to compare with AMS-02 data as observational constraints~\citep{Masi2016}.
 The main propagation parameters were treated as free parameters in the scan using \galprop{} \citep{1998ApJ...509..212S,1998ApJ...493..694M}. 
 The parameters defining the injection spectra, such as spectral indices and the break rigidities, were also treated as free parameters, but their exact values, below $\sim$ 50 GV, depend on the solar modulation. As a matter of fact, the low energy part of the spectra is tuned together with the solar modulation parameters.
 LIS parametrization is mainly constrained  by measurements from Voyager probes~\citep{2016ApJ...831...18C}, at low energy, and AMS-02~\citep{AMS02_2014_PhysRevLett3,AMS02_2014_PhysRevLett2,AMS02_2015_PhysRevLett2,AMS02_2015_PhysRevLett1,AMS02_2016_PhysRevLett1,AMS02_2016_PhysRevLett2,2017PhRvL.119y1101A}, at high energy.
 
 In Fig.~\ref{fig:helmodams02} we summarize the LISs obtained by means of \helmod{}-\galprop{} iterative procedure as reported in \citet[][for protons]{2017ApJ...840..115B},~\citet[][for  helium,  carbon and oxygen nuclei]{2018ApJCO}, and~\citet[][for electrons]{2018ApJElectron}; the corresponding \helmod{} modulated spectra obtained using previous \helmod{} version 3 is shown with AMS-02 measurements reported in \citet[for electron]{AMS02_2014_PhysRevLett2}, \citet[for proton]{AMS02_2015_PhysRevLett1}, and \citet[for  helium,  carbon and oxygen nuclei]{2017PhRvL.119y1101A}. 

\section{The HelMod Heliosphere}
\label{Sect::OuterHeliosphere}
The boundary of the heliosphere, called heliopause (HP), is a contact discontinuity separating the solar cavity -- in which the SW plasma is flowing -- from the interstellar space. It also represents the extreme limit beyond which solar modulation 
does not affect CR flux. Thus, outside HP the truly pristine LIS of GCR spectra could be observed.

After being accelerated in the solar corona \citep{Parker1958} SW adiabatically 
expands in radial direction with supersonic speed. In its journey towards the external regions of the heliosphere SW flow changes its supersonic regime, through the formation of a physical boundary called 
Termination Shock (TS) which, in practice, separates the inner 
part of the heliosphere\footnote{The inner part of the heliosphere -- corresponding to the space region from the Sun up to the TS -- will be indicated as inner heliosphere in the following.} from the outer region\footnote{The outer part of the heliosphere - corresponding to the space region from the TS up to the HP  - will be indicated as outer heliosphere in the following.}, also known as heliosheath (HS).

\par
The heliosphere boundaries (i.e., TS and HP) were extensively discussed in the literature (e.g., \citealp{Parker1961,Parker1963,Axford1972,Holzer1989,Zank1999,Zank2015} and references therein). Recently, fundamental advancements in the knowledge of the outer heliosphere were achieved by means of both Voyager probes, which provided on-site observations of TS and HP positions, SW plasma and magnetic field properties (e.g., see \citealp{Richardson2008,Richardson2013,DatiV2,Burlaga2016,OMNIweb}). In Parker's model of the heliosphere \citep{Parker1961,Parker1963}, such pieces of
information can be exploited for allowing us to estimate the time dependence of both TS and HP positions -- i.e., those currently used in \helmod{} --, as discussed in Sect.~\ref{app1}. For instance, the predicted TS values are in good agreement with those observed: for Voyager~1 (Voyager~2) the detected TS position is 93.8~AU (83.6~AU) and the predicted is 91.8~AU (86.3~AU), i.e. within 3~AU; and, using the HP position observed by Voyager~1, the predicted value is 120.7~AU at the time of the Voyager~2 HP crossing which occurred at 119~AU. Furthermore, the interstellar magnetic field strength measured by Voyager~1 is $(0.48\pm0.04)$nT \citep{Burlaga2016}. In the context of the Parker model, at the TS positions observed by Voyager~1 and Voyager~2, such a magnitude requires that the value of the dimensionless stagnation pressure \citep{Parker1963} approaches its maximum value, i.e., the one allowed for a spherical diamagnetic solar cavity (see discussion in Sect.~\ref{app1}). It should be remarked that, from the measurements of energetic neutral atoms by IBEX \citep{McComas2009} and Cassini \citep{Krimigis2009},  \cite{Dialynas2017} strongly suggested a diamagnetic bubble-like heliosphere\footnote{Under the condition that there is no interstellar magnetic field, \cite{Parker1961,Parker1963} discussed the case of a steady subsonic interstellar wind leading to a comet-like shape \citep[see, e.g.,][and references therein]{Axford1972,Holzer1989,Zank1999,Zank2015}.} with negligibly small tail-like features (see also \citealp{Drake2015,Opher2015,Opher2017}).

\par
In HelMod model, so far the inner region of the heliosphere was described as an \textit{effective heliosphere}~\citep{Bobik2011ApJ} with a radius -- i.e., an effective TS distance~--
of 100AU. Solar modulation was, then, treated by subdividing it in 15 radially equally-spaced regions. Each {\textit{i}-th} region traveled by CR particles is characterized by 
heliospheric parameters evaluated at \textit{i}-Carrington rotations back-in-time,
corresponding to the time needed by SW for reaching it~\citep{Bobik2011ApJ,BoschiniEtAl2017AdvSR}. 
The actual dimensions of the heliosphere are accounted for by scaling the position of TS by the time dependent values obtained as described in Sect.~\ref{app1}.
The HS is included as a single additional region on top of the 15 inner regions with its actual size as calculated in Sect.~\ref{app1}.
As discussed in Sect.~\ref{app1} the TS and HP latitudinal profile in the nose region 
was determined taking into account the latitudinal variation of the ISM ram pressure component of the total pressure on the HP surface. In the anti-nose region only the magnetic and plasma pressure were accounted for. As a consequence, an asymmetry in the direction of the ISM flow is introduced (see Fig.~\ref{fig:bubblelike} in Section~\ref{app1}). 
Finally a cylindrical symmetry with respect to the axis along the sun-nose direction was applied to have full three-dimensional heliospheric boundaries.

\section{Heliospheric boundaries in \helmod{}: time dependent Termination Shock and Heliopause}
\label{app1}


The analytical modelization of the heliosphere dates back to 
\cite{Parker1961} \citep[see also discussions in][]{Parker1963}. 
In the framework of Parker's model the position of the termination 
shock (TS) is obtained from hydro-dynamical considerations. The main
hypothesis is that the heliopause (HP) is a contact discontinuity \citep[e.g., see][]{suess1990} separating the region in which the propagation of the solar wind (SW) dominates from the interstellar medium\footnote{For instance, after November 5 2018 the plasma instrument on board of Voyager~2 has observed no SW flow in the environment around the spacecraft and this can be possibly considered as an experimental observation that the spacecraft entered into the interstellar space \citep{VOY2Announce}. } (ISM).
%
In such a model, the SW flows along streamlines from the inner part of
the heliosphere, passing through the TS and traversing the heliosheath
(HS) up to the stagnation point\footnote{In fluid dynamics, a stagnation point is a point in a flow field where the local velocity of the fluid is zero.} (with $P_{\rm ISM}$ as stagnation pressure) on the HP in a one-dimensional radial approximation. Assuming a spherical heliocentric geometry, the SW is treated as an ideal gas\footnote{For a mono-atomic gas with three degrees of freedom, the specific heat ratio or adiabatic index is $\gamma = 5/3$.} with adiabatic index $\gamma$
expanding steadily, radially and adiabatically towards the TS.
Before reaching the TS, the largely dominant contribution to the total SW
pressure\footnote{The interplanetary magnetic field pressure and thermal pressure can be neglected. In fact, their intensities were estimated to be on average two orders of magnitude smaller than the ram pressure  from on-site spacecraft measurements \citep{OMNIweb}.} is provided by the ram pressure
$p_{ram}=\rho u^2$, where $\rho$ is the plasma density 
and $u$ the SW speed. Moreover, for a constant SW speed up to the TS
\citep{Parker1958} and from the mass conservation, one finds that the
plasma density
and the pressure $p$ scale with distance as
\begin{align}
\rho(R)&=\rho_{obs}(R_{obs}/R)^2, \label{RscaleRho} 
\end{align}
where the subscript $obs$ refers to the quantities measured by an observer located at the heliocentric distance $R_{obs}$. 

\par
As discussed in Parker (1961), before reaching the stagnation point, 
the SW must go through a shock transition: SW plasma abruptly slows 
down and is compressed, so that density
increases\footnote{The magnetic field after TS increases by about a factor 2 \citep[e.g. see][]{V2_2008Natur}. However, the magnetic pressure is still negligible.}. For a shock occurring in a plane
perpendicular to the direction of flow\footnote{For the SW propagation, the plane is perpendicular to the radial direction.} (normal shock) -- as discussed for the SW shock by \cite{Parker1961,Parker1963} --, the hydrodynamical quantities in the neighborhood of the shock are connected by the Rankine-Hugoniot relations (e.g. see \textsection85 chapter~9 in \citealp{LL1959}) which in the {\it strong} shock limit ({\it i.e.}, for high Mach numbers) allow one to provide, for instance, the relationship between particle density, SW velocity and ram pressure at TS, immediately before and immediately after the shock occurrence:
\begin{eqnarray}
p_{ram\,2TS} &=&\rho_{2TS}u_{2TS}^2 \label{RH.pres2}\\
&=&\frac{\gamma-1}{\gamma+1} p_{ram\,1TS}; \label{RH.pres}
\end{eqnarray}
in the above expressions (and in the following) subscript $1TS$ ($2TS$) refers to quantities just before the TS occurs (just after the TS has occurred). 

\par
The Parker hypothesis of {\it strong} shock proved to be a fairly good
approximation also in light of the on-site plasma measurements by Voyager~2 probe.
In fact, the Mach number can be estimated using temperature
and SW speed; one finds that its value is about 9 \citep{Richardson2008} just before the TS (i.e., at 83.6~AU) using the data from Voyager~2 \citep{OMNIweb}.
Furthermore, using the same source of data,
the mean ram pressure calculated in the 2~AU before reaching the TS zone\footnote{We assume that the TS region extends for 1~AU before and after the TS position determined at 83.6~AU by Voyager~2.} located at 83.6~AU is 
$3.26\times10^{-4}$nPa,
while in the 2~AU after the TS zone is $0.83\times10^{-4}$nPa.
Therefore their ratio
is about 3.9 and it is in agreement with the value 4 
-- expected for a strong shock of a monoatomic gas (e.g., see Eq.~\eqref{RH.pres}) -- which, in the Parker model, 
provides the ratio of $1/7{\sim}14.3\%$ (e.g., see~\citealp{Parker1961}) between kinetic pressure\footnote{The kinetic pressure corresponds to $p_{\rm kin}=\frac{1}{2}\rho u^2$.}
and the stagnation pressure\footnote{For Voyager~2 at the TS the ratio between kinetic pressure $\left(\frac{1}{2} 0.83\times10^{-4}{\rm nPa}\right)$
	and the interstellar pressure ($2.62\times10^{-4}$nPa, as discussed later) is about $15.8\%$.	In general, after the shock, the thermal pressure due to ionized and neutral particles (e.g., see discussion in \citealp{Richardson2008}) becomes dominant with a value determined by $P_{\rm ISM}$, as well as the ratio between kinetic and thermal pressure.}.
\begin{figure}
	\begin{center}
		\includegraphics[width=1\linewidth]{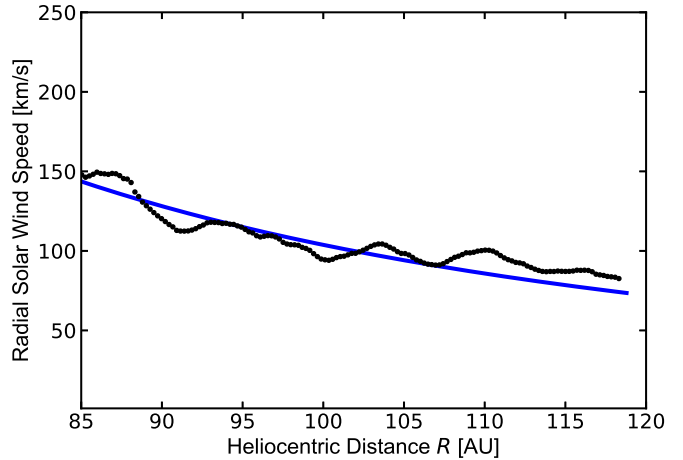}
	\end{center}
	\caption{Radial component of the SW
		speed in the HS ($V_{{\rm sw},2r}$) as function of the heliocentric distance $R$ downstream the Voyager~2 TS: the data are obtained with a coordinate transformation (e.g., see \citealp{Burlaga1984}) from the data
		in \cite{OMNIweb}; the solid line is the $1/R^2$ behavior (see text).}
	\label{fig:SWspeed}
\end{figure}

\begin{figure} 
	\centering
	\includegraphics[width=1.\linewidth]{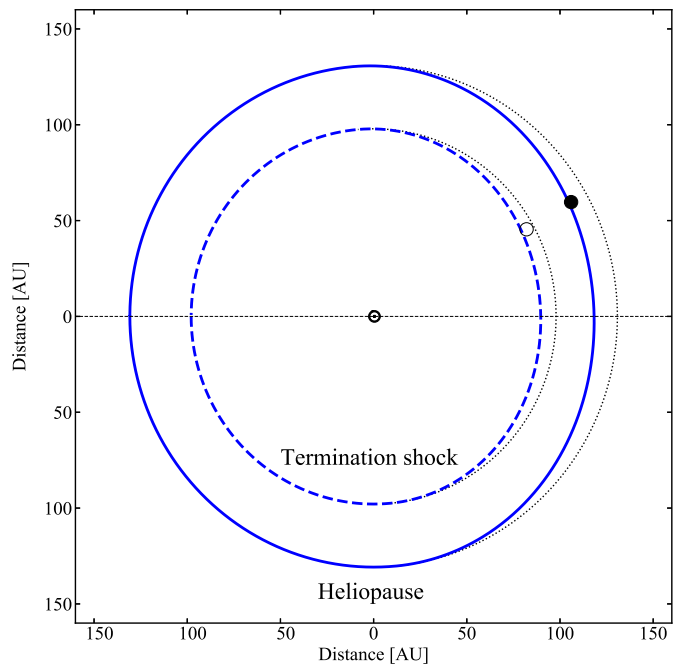}
	\caption{Meridian section of the heliosphere as described in Eq.~\eqref{HelmodBoundaries}. The dashed (solid) line shows the latitudinal profile of the TS (HP) at the time and HCI latitude of the Voyager~1 TS (HP) crossing. The open (full) circle denotes the instantaneous position of Voyager~1 at TS (HP) crossing. 
	}
	\label{fig:bubblelike}
\end{figure}

In addition, although the SW speed in the HS 
does not exhibit an appreciable dependence on $R$ \citep{Richardson2013}, in the region downstream the TS, its
radial component ($V_{{\rm sw},2R}$) progressively slows down flowing in the HS towards the stagnation point \citep{langner2003,DatiV2}. In Fig.~\ref{fig:SWspeed}, the radial SW speed (full circle) is obtained with a coordinate transformation (e.g., see \citealp{Burlaga1984}) from the data in \cite{OMNIweb}. Such a decrease is compatible with the $1/R^2$ behavior
(the solid line) given by:
\begin{equation}
V_{{\rm sw},2R}(R) = u_{2TS} \left( \frac{R_{TS}}{R} \right)^2,
\label{speed.incompr}
\end{equation}
where $u_{2TS}=150.7$~km/s (the SW speed  average value calculated in the 2~AU after reaching the TS zone) and $R_{TS}=83.6$~AU, i.e. the position of the TS observed by Voyager~2.


\par
\begin{table*}
	\centering
	\caption{Physical parameters of the local interstellar medium used in
		this work.}
	\begin{tabular}{clcl}
		\hline
		\multicolumn{2}{c}{Parameter}&Value&Reference\\
		\hline
		\multirow{ 2}{*}{$v_{\rm ISM}$}&\multirow{ 2}{*}{Speed} & \multirow{ 2}{*}{$26$\,km/s$^{-1}$}& \cite{Moebius2004}\\
		&&& \cite{Bzowski2015} \\
		$n_{\rm ISM}$ & Proton number density & $0.07$\,cm$^{-3}$ & \cite{Slavin2008} \\
		\multirow{ 2}{*}{$T_{\rm ISM}$} &\multirow{2}{*}{Temperature} & \multirow{2}{*}{$7000$\,K}
		& \cite{Moebius2004}\\
		&&&\cite{Bzowski2015} \\
		$B_{\rm ISM}$& ISMF strength & 0.50\,nT & (see discussion in the text)\\ 
		\hline
	\end{tabular}
	\label{tab:ISM}
\end{table*}
%
In the Parker model, the TS position is determined as the distance ($R_{TS}$) at which the total pressure in the region downstream 
the shock is equal to the ISM stagnation pressure $P_{\rm ISM}$. \cite{Parker1961} derived two expressions for the TS position depending on 
whether the expansion of the shocked SW in the HS occurs through an {\it isentropic} 
({\it i.e.}, reversible adiabatic) or an {\it incompressible} expansion.
Numerically the relative difference between those two approaches is less 
than $0.4\%$ for a monatomic gas;
for an incompressible flow (i.e., with plasma density 
$\rho =$~const) one finds that the expression for $R_{TS}$ is given by:
\begin{equation}
R_{TS} = R_{obs} \left( \frac{\rho_{obs} u_{obs}^2}{P_{\rm ISM}}
\right)^{\frac{1}{2}} \left[ \frac{\gamma+3}{2(\gamma+1)}
\right]^{\frac{1}{2}}.
\label{RTS.incompr}
\end{equation}

\par
It has to be remarked that the $P_{\rm ISM}$ determines the physical conditions for which the shock occurs at $R_{TS}$, i.e., the inner boundary of the HS. In general, in the HS, the hydrodynamical quantities and its boundaries (TS and HP) are related to $P_{\rm ISM}$.
Voyager~1 has provided the first on-site observation of HP position (121.6~AU on Aug 2012) which allows one to determine, for instance, the value of the outward radial ram pressure of the SW at the outer heliospheric boundary.

\par
In the Eq.~\eqref{RTS.incompr}, the SW ram pressure depends on the solar activity, while $P_{\rm ISM}$ (discussed later)
is commonly assumed to be time independent. Therefore, $R_{TS}$
can be determined as function of time using the monthly averages of the
SW plasma parameters measured by various satellites (e.g., Voyager~2, Wind, ACE, Ulysses) along the last 60
years~\citep{OMNIweb,UFAweb}. The mean value of those monthly averages
finally provides the monthly estimate of $R_{TS}$\footnote{On average the standard deviation is about 5.8~AU} (currently used in \helmod{}).
In Eq.~\eqref{RTS.incompr} the values of $\rho_{obs}$ and $u_{obs}$ are found taking into account the time lag ($\Delta t_{TS}$) needed
to the SW to travel from the observation point at $R_{obs}$ to $R_{TS}$, i.e., 
\begin{align}
\Delta t_{TS}&=\int_{R_{obs}}^{R_{TS}} \frac{dR}{V_{\rm sw}(R)}, \label{eq:TimeDelay}\\
&=\frac{R_{TS}-R_{obs}}{u_{obs}},\label{eq:TimeDelayTS}
\end{align}
where Eq.~\eqref{eq:TimeDelayTS} is employed for a constant SW speed\footnote{The typical value of $\Delta t_{TS}$ amounts to ${\sim}1$ year (15 Carrington rotations) for the SW propagating from 1AU to $R_{TS}{\sim}100AU$ with an average speed ${\sim}450$km/s.} equal to $u_{obs}$.
It must be remarked that using the out-of-ecliptic data from
Ulysses the computed SW ram pressure apparently does not show a
latitudinal dependence. In fact, at high latitudes the SW density decrease is almost 
completely compensated by the SW speed increase. The so calculated ram pressures are well compatible with those obtained with the data from the other satellites within the monthly fluctuations. Therefore, $R_{TS}$ does not exhibit observable latitudinal dependencies.

\begin{figure*}[t]
	\begin{minipage}[t]{1\columnwidth}
		\centering
		\includegraphics[width=1\linewidth]{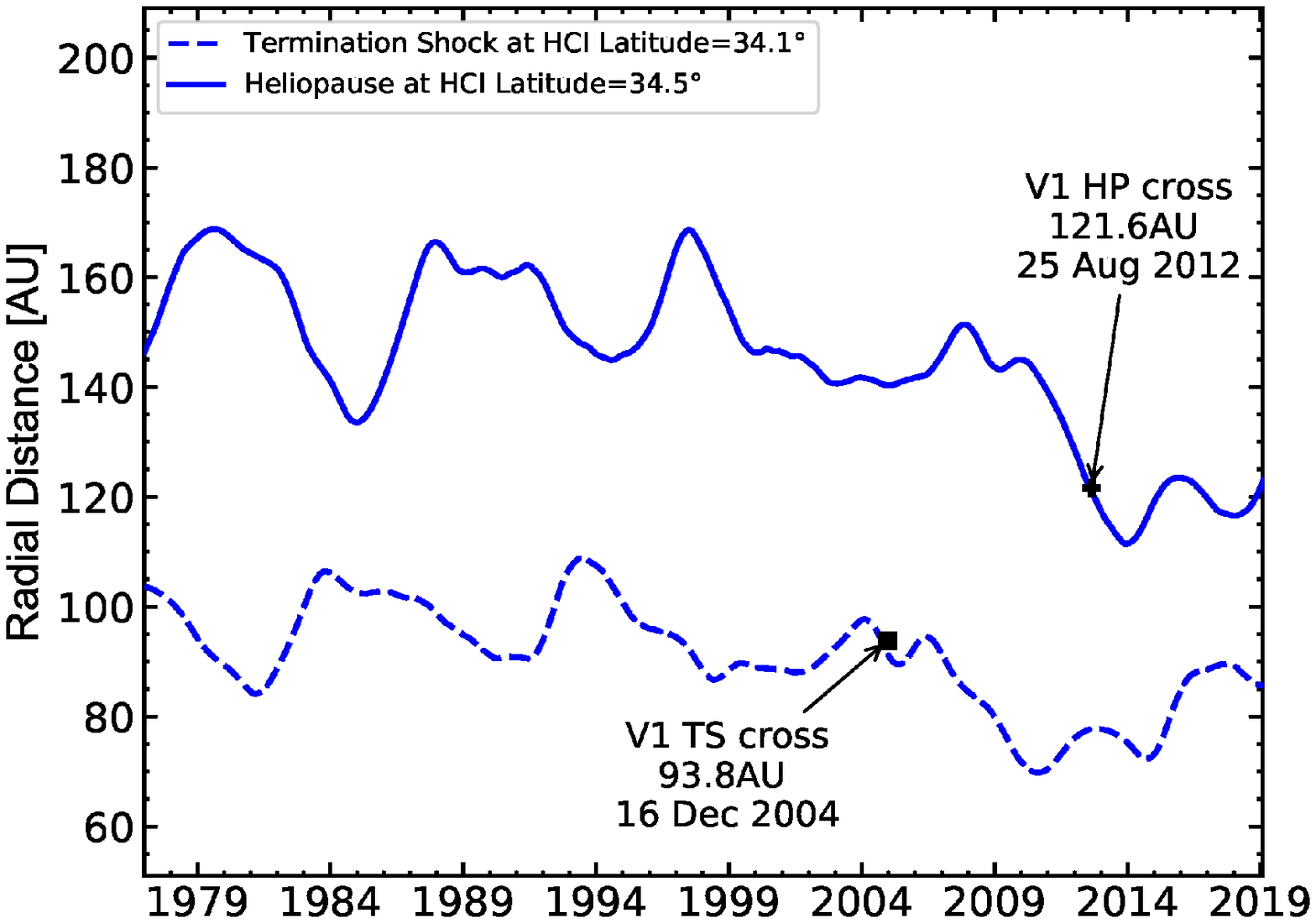}
		\caption{Time variation of monthly averaged position of TS (dashed line) and HP (solid line), calculated from 1977 up to the end of 2018. TS is calculated at HCI-latitude $34.1\degree$, while HP at HCI-latitude $34.5\degree$ to match Voyager~1 trajectory. The predicted TS distance at the time of Voyager~1 crossing is 91.8AU.}
		\label{fig:HPTSdist_V1}
	\end{minipage}
	\hfill
	\begin{minipage}[t]{1\columnwidth}
		\centering
		\includegraphics[width=1\linewidth]{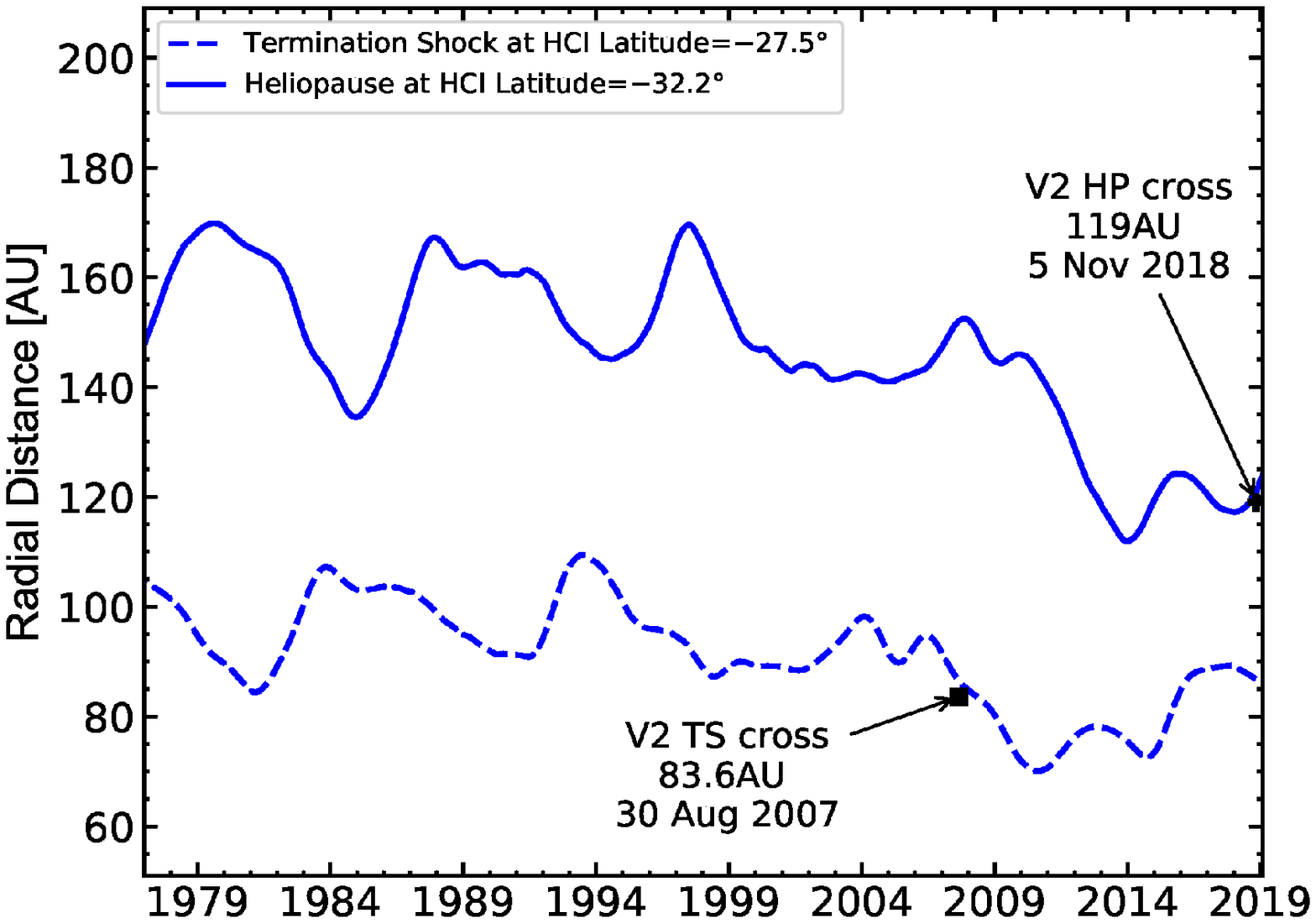}
		\caption{Time variation of monthly averaged position of TS (dashed line) and HP (solid line), calculated from 1977  up to the end of 2018. TS is calculated at heliolatitude $-27.5\degree$, while HP at heliolatitude $-32.2\degree$ to match trajectory of Voyager~2. The predicted TS distance at the	time of Voyager~2 crossing is 86.3AU, while the predicted HP distance at the	time of Voyager~2 crossing is 120.7AU.}
		\label{fig:HPTSdist_V2}
	\end{minipage}
\end{figure*}

\par
\begin{figure*}
	\centering
	\caption{\helmod{} version 4 modulated spectra (red solid line) -- obtained using proton LIS from~\citet{2017ApJ...840..115B} -- compared with measurements at $\sim$0.25 GeV (black points) from instruments on-board of Voyager~1 (upper Panel) and Voyager~2 (lower panel). Voyager's data are from \citep{VoyagerWeb}. Vertical dashed line indicates the TS crossing, while vertical solid line reports the HP crossing of Voyager~1. }
	\label{fig:voyager1}
	\includegraphics[width=0.9\linewidth]{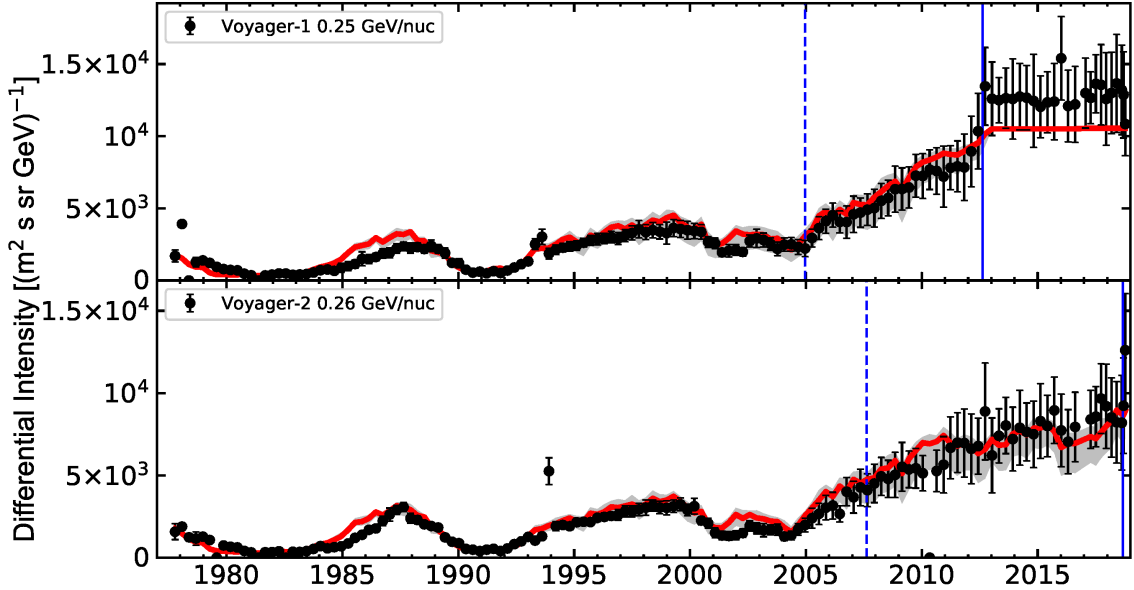}
\end{figure*}

The other quantity appearing in Eq.~\eqref{RTS.incompr} is the stagnation pressure $P_{\rm ISM}$. As already discussed by \cite{Parker1963} (see also \citealp{Parker1961}), to a first approximation $P_{\rm ISM}$ can be
estimated  by adding the interstellar (IS) magnetic field pressure ($p_{\rm mag}$), the kinetic pressure due to IS wind ($p_{\rm kin}$), the thermal pressure of the IS plasma ($p_{\rm th}$), i.e.,
\begin{equation}
P_{\rm ISM}=p_{\rm mag}+p_{\rm kin}+p_{\rm th}.
\label{TotISM}
\end{equation}
The values for $p_{\rm kin}$, $p_{\rm th}$ and $p_{\rm mag}$ are discussed in the following (e.g. see Eqs.~(\ref{pkin},~\ref{pth},~\ref{pmag_val})).
A lower contribution to $P_{\rm ISM}$ is expected from the CR pressure ($p_{\rm CR}$). 
As already discussed by \cite{Parker1963}, $p_{\rm CR}$ accounts for the difference between the pressure derived from the CR omnidirectional intensity in the IS space outside the HP with respect to that one obtained by the CR omnidirectional intensity immediately inside the HP. The CR pressure is determined as in \cite{IpAxford1985} (see also references therein) integrating the CR particle density $U(T)$
for protons and helium nuclei (these two species constitute the dominant contribution to the overall CR omnidirectional intensity).
Their intensities immediately inside the HP are obtained from the modulated \helmod{} spectrum in the HS region close to the HP (e.g., see Section~\ref{Sec.4.2} and the upper panel of Fig.~\ref{fig:voyager1}). The value found for $p_{\rm CR}$ is about $2.4\times10^{-6}$~nPa, i.e., it is of the order or lower than 1\% of the overall $P_{\rm ISM}$ and can be neglected.
The dominant contribution to the stagnation pressure comes from the
IS magnetic field pressure (e.g., see equation~(9.27) of \citealp{Parker1963}):
\begin{equation}
p_{\rm mag}=\Pi^2\frac{B_{\rm ISM}^2}{2 \mu_0},
\label{pmag}
\end{equation}
where $B_{\rm ISM}$ is the average IS magnetic field magnitude, 
$\mu_0$ is the permeability of free space, and $\Pi^2=2.25$ is the dimensionless stagnation pressure (see \citealp{Parker1963}). In Parker's model, this value is the maximum allowed for $\Pi^2$ parameter, and corresponds to an heliosphere described as a spherical ``diamagnetic'' region of radius $l$ in which the SW is streaming away from it along two opposite channels of radius $c\rightarrow0$ along the IS magnetic field direction (e.g., see Figure 9.3 in \citealp{Parker1963}). In addition, the radius of the boundary between the IS magnetic field and the SW, i.e. the heliopause, overlaps that one of the ``diamagnetic'' region.
%
%
The IS wind pressure ($p_{\rm kin}$) is due to the relative motion of the  heliospheric cavity with respect to the ISM. In the HCI reference system\footnote{In the Heliocentric Inertial (HCI) reference frame the x-axis is directed along the intersection line of the ecliptic plane and solar equatorial plane. The z-axis is directed perpendicular to and northward of the solar equator plane, and the y-axis completes the right-handed set (e.g., see \cite{Burlaga1984,FranzHarper2002,FranzHarper2017web}).
} this flow occurs close to the x-axis direction and determines the heliospheric nose direction\footnote{At $\left[178.3\degree,5.1\degree\right]$ in the HCI system of reference \citep{Bzowski2015}.}; $p_{\rm kin}$ is given by
\begin{eqnarray}
p_{\rm kin}&=&\frac{1}{2} n_{\rm ISM} m_p \left[v_{\rm ISM}\cos(\alpha)\right]^2, \label{pkinalp} \\
&=&0.40\times10^{-4}\cos^2(\alpha)~{\rm nPa},
\label{pkin}
\end{eqnarray}
where $n_{\rm ISM}$ (see Table~\ref{tab:ISM}) is the proton number density\footnote{The electron contribution to the overall density is negligible because of their small mass with respect to that of the protons.}, $m_p$ is the proton mass, $v_{\rm ISM}$ (see Table~\ref{tab:ISM}) the relative IS wind speed and, finally,  $\alpha$ is the difference between the HCI heliolatitude of the nose and the heliolatitude of the point at which the pressure is calculated. In Eq.~\eqref{pkinalp}, $v_{\rm ISM}\cos(\alpha)$ is the IS velocity component normal to the HP boundary.
A further contribution comes from the thermal pressure of the interstellar plasma ($p_{\rm th}$):
\begin{eqnarray}
p_{\rm th}&=&2\,n_{\rm ISM} k_B T_{\rm ISM}, \nonumber\\
&=&0.14\times10^{-4}~{\rm nPa},
\label{pth}
\end{eqnarray}
where $k_B$ is the Boltzmann constant, $T_{\rm ISM}$ (see Table~\ref{tab:ISM}) is the ISM temperature and $n_{\rm ISM}$ is its number density; the factor 2 accounts for the equal amount of protons and electrons.

\par
The first on-site observation of $R_{TS}$ at 93.8~AU
was provided by Voyager~1 on 16 December 2004. We have to remark that, since $R_{TS}$ depends on the actual value of $P_{\rm ISM}$ (Eq.~\eqref{TotISM}) whose dominant term is $p_{\rm mag}$ (Eq.~\eqref{pmag}), at the date of Voyager~1 crossing we can estimate
the value of $p_{\rm mag}$, then that of $B_{\rm ISM}$. In fact, $B_{\rm ISM}$ can be derived from Eq.~\eqref{RTS.incompr} introducing 
$R_{obs}$, $\rho_{obs}$ and $u_{obs}$ -- obtained back on time (e.g. see Eq.~\eqref{eq:TimeDelayTS}) from Voyager~2, Wind, ACE, Ulysses (also NASA-OMNIweb) --, the thermal pressure $p_{\rm th}=0.14\times10^{-4}$~nPa (Eq.~\eqref{pth}) and 
the kinetic pressure $p_{\rm kin}=0.30\times10^{-4}~{\rm nPa}$, i.e. calculated at HCI-latitude\footnote{It is worth to mention that the Voyager~1 spacecraft is actually traveling in a direction close to the one of the interstellar magnetic field \citep{Frisch2015,Zirnstein2016}.} of $34.1\degree$ (Eq.~\eqref{pkin}). For $\Pi^2=2.25$ (i.e., that allowed for a spherical diamagnetic solar cavity) the average magnetic field needed to get $R_{TS}=93.8$~AU on 16 December 2004 is $B_{\rm ISM}=(48.6\pm2.0)\times10^{-2}$~nT, well in agreement with the value measured by Voyager~1 in the IS space\footnote{It should be remarked that the consistency between the calculated and observed TS position for Voyager~1 requires, in turn, the maximum value of the dimensionless stagnation pressure $\Pi^2$.}, i.e., $(48.0\pm4.0)\times10^{-2}$~nT \citep{Burlaga2016}.
Using the same procedure, for Voyager~2 which crossed the TS at $R_{TS}=83.6$~AU on 30 August 2007, the kinetic pressure $p_{\rm kin}=0.25\times10^{-4}~{\rm nPa}$, i.e. calculated at HCI-latitude of $-27.5\degree$, for $\Pi^2=2.25$ the average magnetic field needed to get $R_{TS}=83.6$~AU on 30 August 2007 is $B_{\rm ISM}=(51.9 \pm 1.3)\times10^{-2}$~nT. As discussed above, $P_{\rm ISM}$ depends on HCI-latitude $\alpha$ (e.g. see Eqs.~\eqref{TotISM} and \eqref{pkin}), $R_{TS}$ calculated from Eq.~\eqref{RTS.incompr} has to depend, in turn, on $\alpha$ (e.g., see Fig.~\ref{fig:bubblelike}), i.e.,
\begin{flalign}
&R_{TS}(\alpha)= &\label{HelmodBoundaries}\\
&\begin{cases}
R_{TS}(90\degree)
-\left[R_{TS}(90\degree)-R_{TS}(0\degree) \right]\cos^2(\alpha), ~\text{if}\,|\alpha|<90\degree,\\ 
R_{TS}(90\degree)  \hfill \text{otherwise}.
\end{cases}\nonumber&
\end{flalign}
In the nose direction the IS wind affects the extension of the TS by a factor\footnote{This small compression is not present in \cite{Parker1963}, because he discussed the case of a diamagnetic heliospheric cavity in absence of the IS wind.} 
not exceeding 10\%. 

\par
In the present \helmod{} model the $R_{TS}$ positions are obtained from Eq.~\eqref{RTS.incompr} using the mean of the two so derived $B_{\rm ISM}$ values (see Table~\ref{tab:ISM}); the corresponding $p_{\rm mag}$ to be used in Eq.~\eqref{TotISM} becomes:
\begin{equation}
p_{\rm mag}=2.24\times10^{-4}~{\rm nPa}.
\label{pmag_val}
\end{equation}
In Fig.~\ref{fig:HPTSdist_V1} (Fig.~\ref{fig:HPTSdist_V2}) the time variation of the 
positions of the TS (dashed line)  from 1977 to up to the end of 2018 is shown at $34.1\degree$ ($-27.5\degree$). By inspecting the two figures, one can remark that the predicted values are in good agreement with those observed: for Voyager~1 (2) the observed TS position is 93.8~AU (83.6~AU) and the predicted is 91.8~AU (86.3~AU), i.e. within 3~AU.

\begin{figure*}
	\centering
	\caption{ \helmod{} version 4 solutions at 2 GV (red solid
          line) for protons (top panel), helium nuclei (central
          panel), and electrons (bottom panel) obtained using the
          corresponding LISs
          from~\citet{2017ApJ...840..115B},~\citet{2018ApJCO},
          and~\citet{2018ApJElectron}, respectively.  In the same
          plot, experimental data, at the nearest rigidity bin, from
          EPHIN, BESS, PAMELA and AMS-02 are reported. }
	\includegraphics[width=0.9\linewidth]{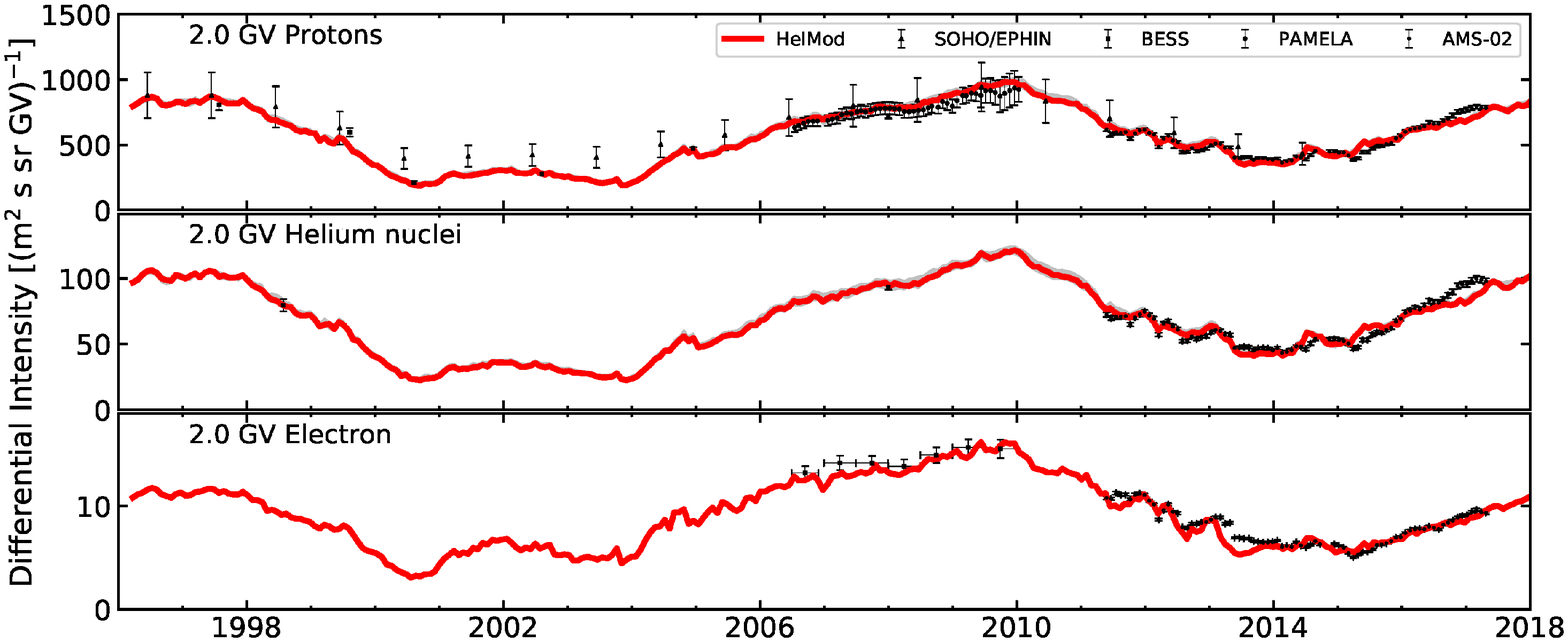}
	
	\label{fig:FullPeriod_P_He_02}
\end{figure*} 

\begin{figure*}
	\centering
	\caption[AMS-02]{Modulated spectra observed by AMS-02 (points), for protons~\citep{AMS02_2015_PhysRevLett1}, electrons~\citep[]{AMS02_2014_PhysRevLett2} and nuclei, i.e., helium, carbon and oxygen~\citep{2017PhRvL.119y1101A}, compared with \helmod{}-\galprop{} LISs (dashed lines, already shown in Fig.~\ref{fig:helmodams02}) from \citet{2017ApJ...840..115B},~\citet{2018ApJCO}, and~\citet{2018ApJElectron} along with \helmod{} version 4 modulated spectra (red solid line). On the left panel, differential intensities are reported. On the right panel, the relative difference of experimental data with \helmod{} version 4 simulations and LISs are shown.
	}
	\includegraphics[width=1\linewidth]{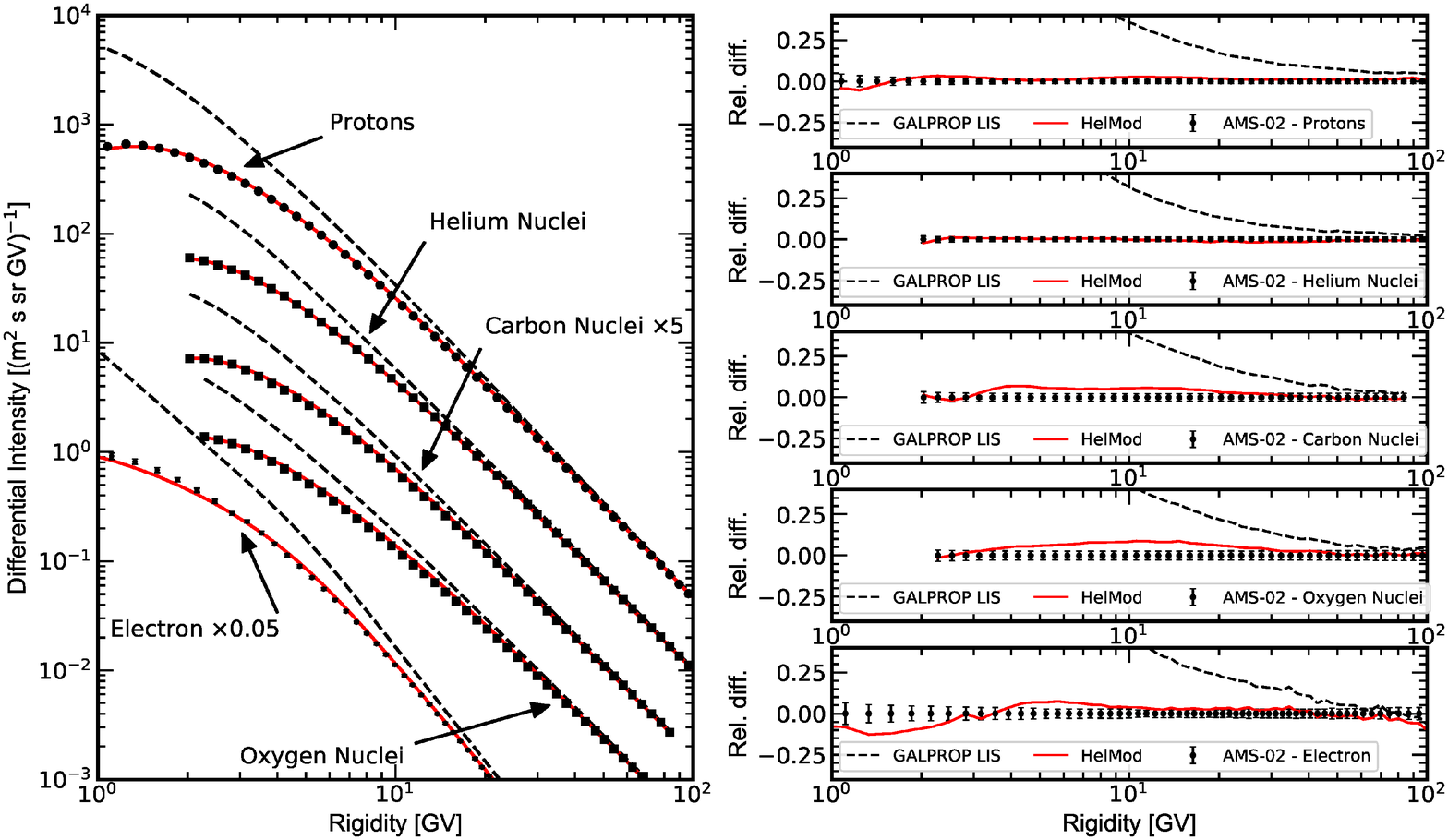}
	\label{fig:AMS02}
\end{figure*}

\par
As already discussed, the ram pressure in the radial direction at the HP (as well as that at the TS) depends on $P_{\rm ISM}$. For the motion of an incompressible fluid reaching the HP at the position $R_{HP}$, one finds that, independently of the HCI latitude, the ratio ($T_H$) of the radial ram pressure at the TS with respect to that at the HP using Eqs.~\eqref{RH.pres2} and \eqref{speed.incompr} is
\begin{equation}
T_H
=\left(\frac{R_{HP}}{R'_{TS}}\right)^4,
\label{eq:ratio}
\end{equation}
where $R'_{TS}$ is the TS position back on time\footnote{The typical value of $\Delta t_{HP}$ amounts to ${\sim}4$ years.} by $\Delta t_{HP}$, i.e., that at which the SW stream was leaving the TS with speed $u'_{2TS}$. In the current \helmod{} model, $R'_{TS}$ are obtained from Eq.~\eqref{RTS.incompr} using the monthly averages of the SW plasma parameters measured by various satellites, as previously discussed. Thus the HP boundaries\footnote{In the nose direction the interstellar wind affects the extension of the HP by a factor not exceeding 10\%.} are described by Eq.~\eqref{HelmodBoundaries} by replacing the $R'_{TS}(\alpha)$ with $R_{HP}(\alpha)$ once the ratio $T_H$ is determined using Voyager~1 observations (e.g., see Fig.~\ref{fig:bubblelike}). It should be remarked that, from the measurements of energetic neutral atoms by IBEX \citep{McComas2009} and Cassini \citep{Krimigis2009},  \cite{Dialynas2017} strongly suggested a diamagnetic bubble-like heliosphere\footnote{Under the condition that there is no interstellar magnetic field, \cite{Parker1961,Parker1963} discussed the case of a steady subsonic interstellar wind
	leading to a comet-like shape \citep[see, e.g.,][and references
	therein]{Axford1972,Holzer1989,Zank1999,Zank2015}.} with few substantial tail-like features (see also \citealp{Drake2015,Opher2015,Opher2017}).
$\Delta t_{HP}$ can be computed using Eq.~\eqref{eq:ratio} and by introducing the the appropriate quantities in Eq.~\eqref{eq:TimeDelay}; finally, one gets
\begin{equation}
\Delta t_{HP}
=\frac{R'_{TS}}{3 u'_{2TS}}\left( \sqrt[4]{T_H^3}
-1 \right).  
\label{eq:HPTimeDelay}
\end{equation}

\par
Voyager~1 provided the first on-site measurement of $R_{HP}$ (121.6~AU on 25 August 2012). The estimation of the corresponding $R'_{TS}$ can be obtained using SW speed measurements from Voyager~2, Wind, ACE, Ulysses (also NASA-OMNIweb). In fact, for any position on TS calculated from Eq.~\eqref{RTS.incompr}, one can determine that particular value $R'_{TS}$ at which the SW has a speed $u'_{2TS}$ such that it reaches the HP position at the exact crossing date of Voyager~1. The average of the so obtained $R'_{TS}$ values is $77.2\pm2.6$~AU. Thus, one finds that numerically Eq.~\eqref{eq:ratio} can be re-written as
\[\frac{R_{HP}}{R'_{TS}}=\sqrt[4]{T_H}=1.58 \pm 0.05.\] 

\par
In Fig.~\ref{fig:HPTSdist_V1} (Fig.~\ref{fig:HPTSdist_V2}) the time variation of the average positions of the HP (solid line) from 1977 to up to the end of 2018 is shown at $34.5\degree$ ($-32.2\degree$). By inspecting the two figures, one can remark that the predicted value of 120.7~AU at the time of the Voyager~2 HP crossing\footnote{
	Due to the slow speed of the spacecraft, we cannot exclude a further crossing of  HP boundary in the future.} (5 November 2018) is in agreement with that observed, i.e., 119~AU \citep{VOY2Announce}.

\section{Solar Modulation in the outer heliosphere}
\label{Sec.4.2}
As discussed in~\citet{2016JPhCS.767a2014K}, the diffusion and drift motion of CRs in the HS should depend on the structure of the HMF beyond the TS. Inside the HS, although several authors propose more complex models \citep[see, e.g.,][]{2015PhPl...22i1501Z,2016JPhCS.767a2014K}, to a first approximation, currently in \helmod{} we implemented a spherical symmetric propagation description depending on a scalar diffusion coefficient $k_{\rm hs}=2.5\times 10^{-5} \cdot \beta \cdot P $ AU$^2$ s$^{-1}$. \citet{SchererEtAl2011} already discussed that in the HS the plasma flow is to a good approximation incompressible (see also discussion in Sect.~\ref{app1}) and, therefore, divergence-free, implying vanishing adiabatic energy changes. Under such a condition Parker equation reduced to (e.g. see \citealt{SchererEtAl2011,DellaTorre2016_OneD}):
\begin{equation}\label{EQ::ParkerEQ_sph_1D_U}
\frac{\partial U}{\partial t}=   \frac{1}{R^2}\frac{\partial  } {\partial R}\left(R^2 k_{\rm hs}\frac{\partial}{\partial R} U\right) - \frac{1}{R^2}\frac{\partial R^2 V_{{\rm sw},2R} U}{\partial R}. 
\end{equation}
where $V_{{\rm sw},2R}$ is the radial SW speed in the HS (see Eq.~\eqref{speed.incompr}). The diffusion coefficient was tuned in order to reproduce Voyager proton spectra during the journey in the HS (see discussion on Fig.~\ref{fig:voyager1} in Sect. \ref{sect:discussion}).

\par
In order to account for the strong modulation effect observed by Voyager 1 in 2012 before the HP crossing~\citep[see, e.g.,][and references therein]{2015PhPl...22i1501Z}, the diffusion coefficient must be reduced by a factor 50 in the outermost layer, 1--2 AU thick, thus  allowing the creation of a diffusion barrier against low energy CRs propagation.
As shown in Fig.~\ref{fig:voyager1} (discussed in Sect. \ref{sect:discussion}), available data of CR in HS, from Voyagers probes, are well reproduced within this simplified scenario. 

\section{Solar Modulation in the inner part of the heliosphere}
\label{Sect::Results}
Modeling solar modulation for a time period covering  more than one solar cycle represents a challenge due to the large variability of the interplanetary environment not only from the solar minimum up to the solar maximum but also from one cycle to the next one.
Nevertheless, particle modulation occurring during solar minima is well described by 
the transport model -- including magnetic drift -- presented in Sect.~\ref{Sect:Model}. Other approaches  leading to similar results can be found, e.g., in \citet{2013JAdR....4..259P} and reference there in (see also, \citealp{2004ApJ...603..744F,2019ApJ...871..253C}). 
\helmod{} code~\citep{GervasiEtAl1999,Bobik2011ApJ,DellaTorre2013AdvAstro,BoschiniEtAl2017AdvSR} is a Monte Carlo numerical code that solves Eq.~\eqref{EQ::FPE} using Stochastic Differential Equations, in a backward-in-time approach described, e.g., in~\citet{DellaTorre2016_OneD}.

%

The current version of \helmod{} model treats solar modulation separately in the inner and in the outer heliosphere. 
In the inner heliosphere, confined by the slightly asymmetric TS described in Sect.~\ref{Sect::OuterHeliosphere},  we use the time-dependent propagation model described in appendix A.1 of~\citet{BoschiniEtAl2017AdvSR}. 
The current model exhibits a weak longitudinal dependence that originates from the presence of the asymmetric TS. Nevertheless, at 1AU the difference of computed flux between nose and tail directions of the heliosphere is less than the numerical method uncertainties ($<$ a few \%).

\helmod{} was tuned on proton flux in both low and high solar activity periods, allowing to reproduce intensity variation along the complete cycle duration. 
The qualitative agreement among experimental data and simulated spectra can be appreciated in Fig.~\ref{fig:FullPeriod_P_He_02}, where \helmod{} results are compared to protons, helium nuclei and electrons differential intensities at 2 GV from the begin of 1996 up to the end of 2017. For such a comparison we include data from AMS-02~\citep{2018PhRvL.pHe,2018PhRvL.epos}, PAMELA~\citep{PamelaProt2013,2015ApJ...810..142A}, BESS~\citep{bess_prot} and SOHO/EPHIN \citep{2016SoPh..291..965K}, thus covering the last two solar cycles. In Fig.~\ref{fig:AMS02} representative examples of comparison of \helmod{} 4 simulated spectra with AMS-02 data~\citep{AMS02_2015_PhysRevLett2,AMS02_2015_PhysRevLett1,AMS02_2014_PhysRevLett2,2017PhRvL.119y1101A} are presented\footnote{Calculation for different experimental data-sets can be downloaded using the HelMod Web Modulator (www.helmod.org)}; the agreement, as shown in right panels of  Fig.~\ref{fig:AMS02}, is better than previously obtained and  showed in Figure 11 of \citep{2017ApJ...840..115B}, in Figure 3 of \citep{2018ApJElectron}, in Figure 4 and 5 of \citep{2018ApJCO}, and, finally, reported above in Fig.~\ref{fig:helmodams02}.

\begin{table*}
	\centering
	\caption{\helmod{} parameters for the transition function described in Eq.~\eqref{EQ::TimeDep}.}\label{tab::par1}
	\begin{tabular}{clcccccc}
		{}						&&        {}          &		{}			 & \multicolumn{2}{|c|}{Ascending} &  \multicolumn{2}{|c|}{Descending}\\
		{}						&& $F_{\textrm{min}}$ & $F_{\textrm{max}}$&  \multicolumn{1}{|c}{$\alpha_0$} & \multicolumn{1}{c|}{$s$ } &   $\alpha_0$  &\multicolumn{1}{c|}{$s$ }       \\ 
		\hline \hline
		$P_{0,d}$ 				&& 0.5			 & 4 					& 73 	& 1						 & 63 	& 10 \\ 
		\hline
		$P_{0,NS}$  			&& 0.5			 & SSN$^*$/50				& 73 	& 1						 & 63 	& 10 \\ 
		\hline
		\multirow{ 4}{*}{$K_c$} & $q>0$; $A>0$& 3			 & 1 					& 40 	& 18					 & 53 	& 5 \\
		& $q>0$; $A<0$& 1			 & 1 					& - 	& -					 & - 	& - \\ 
		& $q<0$; $A<0$& 3			 & 1 					& 40 	& 18					 & 53 	& 5 \\
		& $q<0$; $A>0$& 0.7			 & 1 					& 47 	& 5.8					 & 58.4	& 5.8 \\ 
		\hline
		\multirow{ 2}{*}{$g_{low}$}& e$^+$ ;e$^-$& 0.4		 & 0 					& 67 	& 20					 & 45 	& 10 \\
		& p; Ions	& 0.5			 & 0 					& 60 	& 9						 & 45 	& 10 \\
		\hline
		$R_c$	    			&& 4			 & 1 					& 60 	& 9						 & 45 	& 10 \\
		\hline \hline
		\multicolumn{8}{p{10cm}}{\small $^*$ SSN is the smoothed sunspot number from SIDC~\citep{sidc,ssn_v2}.}
	\end{tabular}
\end{table*}

\begin{figure*}
	\centering
	\caption[AMS ratio with time]{The p/He flux ratio as function
          of time for 9 characteristic rigidity bins as reported in
          \citet{2018PhRvL.pHe}. \helmod{} version 4 simulation (solid
          line) -- obtained using the corresponding LISs
          from~\citet{2017ApJ...840..115B},~\citet{2018ApJCO} -- is
          compared with AMS-02 and BESS observation at nearest
          rigidity bin. Vertical bars on the right side represent 5\%
          variation. }
        \includegraphics[width=0.8\linewidth]{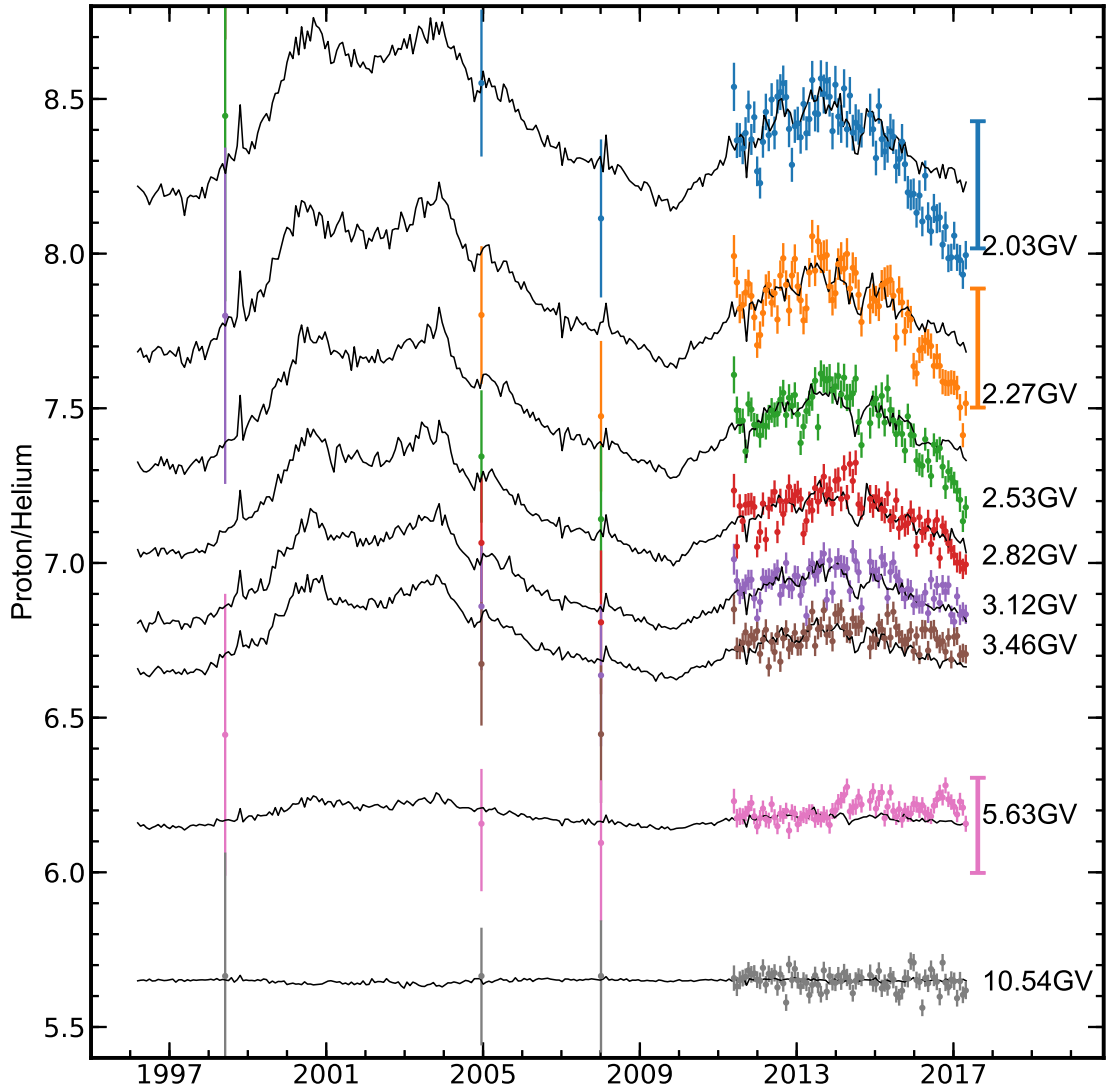}
	\label{fig:pHe_AMS}
\end{figure*}

\subsection{Transition from low to high solar activity}
The high accuracy of newest AMS-02 data, that presented differential intensity of protons, helium nuclei and electron from 2011 to 2017 down to 1 GV~\citep{2018PhRvL.pHe,2018PhRvL.epos}, made possible to study with great details the solar maximum of cycle 24.
It is well known that the effects related to particle drift processes are reduced at solar maximum due to the more chaotic structure of the HMF~\citep[see, e.g.,][]{Minnie2007,Burger2010}. 
This is usually accounted in Eq.~\eqref{EQ::FPE} by means of a partial (or, eventually, complete) suppression of the drift term during the solar maxima~\citep[see, e.g., discussion in ][]{2004ApJ...603..744F}.
Moreover, the presence of turbulences in the interplanetary medium should reduce the global effect of CR drift at very low energy also during solar minimum periods. In literature this effect is usually taken into account by a \textit{drift suppression factor}, $f_s$, that is more relevant at rigidities below 1 GV
\citep[see, e.g., Equation 2 in ][and references therein]{2017ApJ...841..107E}:
\begin{equation}
f_s=\frac{(P/P_{0,d})^2}{1+(P/P_{0,d})^2}.
\end{equation}
 $P_{0,d}$, in GV, is an ad-hoc parameter tuned to achieve the agreement with data.
In the present model this \textit{suppression factor} is applied with a different magnitude both on \textit{regular drift} and \textit{neutral sheet drift}.
The parameters $P_{0,d}$ (for \textit{regular drift} suppression) and $P_{0,NS}$  (for \textit{neutral sheet drift} suppression) were derived by comparison with time dependent spectra measured by AMS-02. A crosscheck of such a parametrization was done using BESS data during solar maximum of solar cycle 23.
Some of these parameters -- i.e. $P_{0,d}$, $P_{0,NS}$, $K_c$, $g_{low}$ and $R_c$ -- change their value with solar activity by means of a transition function from solar minimum to maximum assuming the form:
\begin{equation}\label{EQ::TimeDep}
F(\alpha_t)=\frac{F_{\textrm{min}}+F_{\textrm{max}}}{2}-\frac{F_{\textrm{min}}-F_{\textrm{max}}}{2}\tanh\left[\frac{\alpha_t-\alpha_0}{s}\right].
\end{equation}
Parameters of Eq.~\eqref{EQ::TimeDep} are reported in Table~\ref{tab::par1}; $F_{\textrm{min}}$ represents the parameter value at solar minimum, $F_{\textrm{max}}$ at solar maximum, $\alpha_t$ is the "L"-model tilt angle of neutral sheet as computed by Wilcox Observatory~\citep[][see discussion in \citealt{Bobik2011ApJ,DellaTorre2013AdvAstro,BoschiniEtAl2017AdvSR}]{Hoeksema1995,WSO}, $\alpha_0$ and $s$ are parameters defining the time and the sharpness of the transition.

By inspection of Table~\ref{tab::par1}, it is worth to note that the \textit{neutral sheet drift suppression} at solar maximum depends on the amount of solar disturbances (parametrized by smoothed sunspot number as a proxy): the higher the solar activity, the broader the rigidity range of drift suppression.
Conversely, in the case of \textit{regular drift suppression} factor the rigidity range at maximum is always the same. 


By tuning AMS-02 and PAMELA measurements, i.e., at solar maximum and at solar minimum respectively, we found different values of  $g_{\rm low}$ and $R_c$ parameters to be employed in Eq.~\eqref{EQ::KparActual}. The transition between the two regimes was parametrized by means of  Eq.~\eqref{EQ::TimeDep}. 
Parameters reported in Table~\ref{tab::par1} were crosschecked using BESS and AMS-01 data during the solar minimum between solar cycles 22 and 23.
Finally the parallel component of the diffusion tensor is multiplied by a correction factor ($K_c$) that rescales the absolute value of $K_{||}$ due to drift contribution~\citep{BoschiniEtAl2017AdvSR}. Such a correction is more important during solar minima  when drift processes are relevant. The time evolution of $K_c$ is still parametrized by means of Eq.~\eqref{EQ::TimeDep} using the parameters reported in Table~\ref{tab::par1}.

%
\begin{figure*}
	\centering
	\caption[Ions ratio with time]{Nuclei flux ratios from 1996 to
          2017 for ${\sim}$2~GV determined by means of \helmod{}
          version 4 simulations, using the corresponding LISs
          from~\citet{2018ApJCO}, are reported with solid line. Top
          panel shows carbon over helium ratio, mid panel oxygen over
          helium ratio and low panel carbon over oxygen ratio. Dark
          (light) blue regions represent the 2\% (5\%) variation. }
        \includegraphics[width=0.85\linewidth]{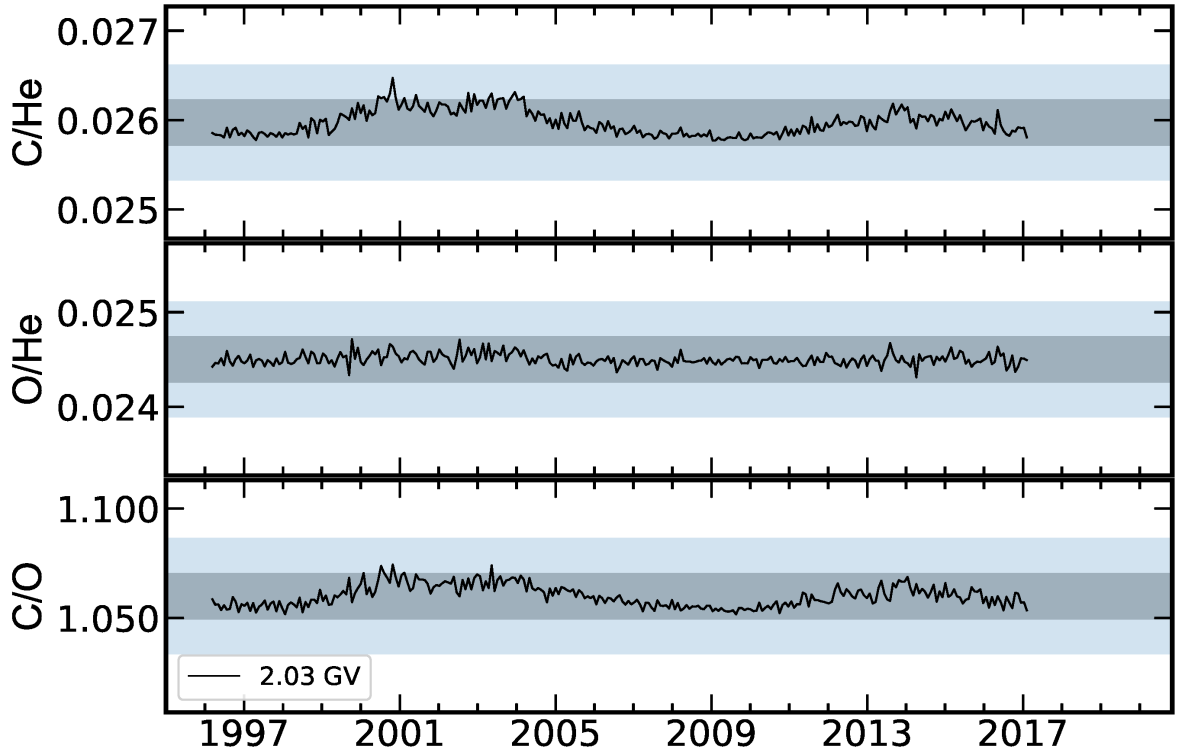}
	\label{fig:ionsRatio}
\end{figure*}

\subsection{Discussion}\label{sect:discussion}

After being tuned using proton flux, \helmod{} model is then applied to evaluate nuclei and electron spectra along the last two solar cycles. In Fig.~\ref{fig:FullPeriod_P_He_02} 
 the \helmod{} simulations for 2 GV  protons, helium nuclei and electrons are shown. 
Typical rms discrepancy between data and simulations is of the order of 7\%. This value becomes 10\% if SOHO data are included.
%
%
Discrepancies greater than these values are related to specific limited periods that need further investigations. 
As indicated in right panel of Fig.~\ref{fig:AMS02}, at 20 GV modulation is still active. For example, the difference between LISs and AMS-02 observed spectra is 15.7\%, 13.2\% and 21.8\% for protons, helium nuclei and electrons respectively
(see also Figure 11 of~\citealt{2017ApJ...840..115B} and Figure 3 of~\citealt{2018ApJElectron}).
This difference is slightly dependent on time, it is almost charge independent, and it is mostly related to diffusion term of Eq.~\eqref{EQ::FPE}.

\helmod{} can be also used to estimate the GCR omni-directional intensities at large distances from Earth and outside the ecliptic plane in comparison with  data from Voyager~\citep[see, e.g.,][]{1986GeoRL..13..781M,CummingsetAl1987,1990SSRv...52..121V,2005IJMPA..20.6727Z} and Ulysses (see \citealt{Heber2011} or \citealt{Heber2006} for recent reviews) probes.  
\helmod{} already demonstrated, also with previous versions, its capability to reproduce qualitatively and quantitatively the latitudinal profile of the GCR intensity as observed in the inner part of heliosphere by the Ulysses
spacecraft. \helmod{} 4.0 simulations confirm results obtained previously \citep[e.g. see][]{DellaTorre2013AdvAstro,BoschiniEtAl2017AdvSR,2017ApJ...840..115B,2018ApJElectron}.
To account the radial dependence observed by Voyager probes,
a systematic investigation of the parallel and perpendicular diffusion coefficients (Sect.~\ref{Sect:Model}) at large heliocentric distances was carried out.
Several radial profiles for $K_{||}$ and $K_\perp$  were tested~\citep[see, for example,][]{1998ApJ...505..244B,2016A&A...591A..18L,2018ApJ...859..107M}.
The best agreement was obtained using the analytical expression in Eq.~\eqref{EQ::KparActual}, implying a larger diffusion coefficient -- and, in turn, less modulation effect -- in the inner part of heliosphere with respect to a simpler radial dependence.  
In Fig.~\ref{fig:voyager1} \helmod{} results are compared with the experimental data from Voyager~1 and Voyager~2 at 0.25 GeV~\citep{VoyagerWeb}.
Current model shows a general good agreement with experimental data except for some specific period, i.e. declining phase in 1987, that needs further investigation.

\par
Recent measurements of AMS-02~\citep{2018PhRvL.pHe} pointed out that the p/He ratio has a long-term dependence on time for rigidities lower than ${\sim}$3 GV.
In Fig.~\ref{fig:pHe_AMS} we reported the p/He ratios obtained from HelMod model from 1996 to 2017 compared with observations made by AMS-02 and BESS~\citep{AMS_helium,AMS01_prot,BESS2007_Abe_2016,2018PhRvL.pHe}. By inspection of HelMod results, one can note that the p/He ratio exhibits a time dependence with a periodicity longer than the duration of the AMS-02 data taking.
For the first time, the high precision data from AMS-02 allowed to highlight the variation of p/He ratio over a long period of time.
In Fig.~\ref{fig:ionsRatio} the \helmod{} predictions are shown to illustrate the time dependence of the flux ratios for carbon over helium, oxygen over helium and carbon over oxygen at ${\sim}$2~GV from 1996 to 2017;  in Fig.~\ref{fig:ionsRatio} the HelMod predictions are shown to illustrate the time dependence of the flux ratios for carbon over helium, oxygen over helium and carbon over oxygen at 2 GV from 1996 to 2017; these ratios are predicted to be almost independent of time also at such a low rigidity.  These simulations indicate that a tiny variation with time -- slightly lower than those presented in Fig.~\ref{fig:pHe_AMS} for the p/He ratio -- is exhibited with differences on average of the order of 2\% (dark blue bands) among solar maxima and solar minima.
%


\section{Conclusions}
\label{Cncl_sect}
In this work, we presented the current version 4 of \helmod{} model which deals with the modulation processes affecting GCRs, during their propagation in the inner and outer regions of the heliosphere. 
\par
In the present code particular attention was paid to deal with high solar activity periods, by comparing our simulations with observations made by AMS-02, and on transitions from/to low solar minima. This was achieved by introducing a \textit{drift suppression factor} and particle diffusion parameters which depend on the level of solar disturbances. In \helmod{}, time-dependent heliospheric parameters were tuned by comparison with the statistically dominant proton spectra, then used to derive the modulated spectra for all GCR species. The solidity of \helmod{} model was demonstrated by its capability of reproducing protons, nuclei and electrons CRs spectra observed during solar cycles 23-24 by several detectors, for instance, PAMELA, BESS and AMS-02.
\par
In the present model the actual dimensions of the heliosphere and its boundaries were taken into account based on Voyager probes observations. The modulation in HS was investigated by means of a 1-D solution, which turns out to be well reproducing the Voyager measurements, for instance, those regarding the intensity of GCRs as function of heliocentric distance for the Voyager~1 (Voyager~2) energy channel of 0.25~GeV/nuc (0.26~GeV/nuc).
In addition, the agreement is also found among the \helmod{} simulations and Voyager intensity measurements in the inner part of the heliosphere. This is an indication of the appropriateness of the modulation mechanisms implemented in the model.

%
%
\section*{Acknowledgements}

This work is supported by ASI (Agenzia Spaziale Italiana) under contract ASI-INFN I/002/13/0 and ESA (European Space Agency) contract 4000116146/16/NL/HK.
\par
We acknowledge the NMDB database (www.nmdb.eu), supported under the European Union's FP7 programme (contract no. 213007) for providing data.~The data from McMurdo were provided by the University of Delaware with support from the U.S. National Science Foundation under grant ANT-0739620.~Finally, we acknowledge the use of NASA/GSFC's Space Physics Data Facility's OMNIWeb service, and OMNI data.
\par
We wish to specially thank Pavol Bobik, Giuliano Boella, Davide Grandi, Karel Kudela, Simonetta Pensotti, Marian Putis, Davide Rozza, Mauro Tacconi and Mario Zannoni for their support to \helmod{} code and suggestions.
%


\section*{References}

\bibliography{mybibfile}

%
%

\clearpage


\end{document}